\documentclass[aps,pra,twocolumn,amsmath,amssymb,superscriptaddress,reprint]{revtex4-2}

\usepackage[utf8]{inputenc}
\usepackage{graphicx}
\usepackage[colorlinks=true,citecolor=blue]{hyperref}
\usepackage{units}
\usepackage{bm}

\newcommand{\tL}{\text{L}}

\newcommand{\tH}{\text{H}}
\newcommand{\tE}{\text{E}}
\newcommand{\tG}{\text{G}}
\newcommand{\tf}{\text{f}}
\newcommand{\ts}{\text{s}}
\newcommand{\circled}[1]{{\raisebox{0.4pt}{\large \textcircled{\small #1}}} }

\begin{document}
\title{Statistical analysis of spin switching in coupled spin-crossover molecules}
\author{Philipp Stegmann}\email{psteg@mit.edu}
\affiliation{Theoretische Physik, Universit\"at Duisburg-Essen and CENIDE, D-47048 Duisburg, Germany}
\affiliation{Department of Chemistry, Massachusetts Institute of Technology, Cambridge, Massachusetts 02139, USA}
\author{Alex Gee}
\affiliation{Department of Materials, University of Oxford, 16 Parks Road, Oxford, OX1 3PH, UK}
\author{Neil T. Kemp}
\affiliation{Department of Physics and Mathematics, University of Hull, HU6 7RX, United Kingdom}
\affiliation{School of Physics and Astronomy, University of Nottingham, Nottingham, NG7 2RD, United Kingdom}
\author{Jürgen König}\email{koenig@thp.uni-due.de}
\affiliation{Theoretische Physik, Universit\"at Duisburg-Essen and CENIDE, D-47048 Duisburg, Germany}
\date{\today}

\begin{abstract}
We study the switching behavior of two spin-crossover molecules residing in a nanojunction device consisting of two closely spaced gold electrodes.
The spin states are monitored through a real-time measurement of the resistance of the junction.
A statistical analysis of the resistance values, the occupation probabilities, and the lifetimes of the respective spin states shows that the two spin-crossover molecules are coupled to each other.
We extract the parameters for a minimal model describing the two coupled spin-crossover molecules.
Finally, we use the time dependence of factorial cumulants to demonstrate that the measured data indicates the presence of interactions between the two spin-crossover molecules.
\end{abstract}

\maketitle

\section{\label{sec:intro}Introduction}

Spin-crossover molecules (SCO) are promising bi-stable magnetic switches with potential application in molecular spintronics.
Under the influence of an external stimulus, such as change of pressure or temperature, spin-crossover compounds can undergo a transition between a high-spin and a low-spin configuration \cite{Cambi_1931}. 
The collective switching behavior of bulk samples or films has been well studied \cite{Gutlich_2000,Kipgen_2021,Shi_2009}.

In order to use spin-crossover-based magnetic switches in applications, e.g., as building blocks for data storage, miniaturization of the device towards the nanoscale is required, with small clusters or even individual molecules being contacted by electrodes.
This may fundamentally change the switching behavior since, first, the interaction between individual molecules strongly depends on whether they reside in a crystalline bulk sample or are only loosely connected to each other and, second, the interaction between molecules and substrate becomes of dominating importance.
Instead of acting collectively, individual molecules will switch individually, with switching properties determined by details of the device \cite{Gruber_2020}.
This requires a careful and profound experimental and theoretical study of the switching behavior of individual molecules.

Single-molecule switches based on spin-crossover molecules have been realized in a number of experiments. Among these, scanning tunnelling microscopy (STM) and other surface probe based techniques have been invaluable tools in revealing the surface dynamics and electrical properties of individual and small clusters of SCO complexes, whilst also providing proof of concept that the transition could be used as a conductance switching mechanism \cite{Alam_2010,Naggert_2011,Palamarciuc_2011,Miyamachi_2012,Gopakumar_2012,Gopakumar_2013,Pronschinske_2013,Gruber_2014,Aragones_2016,Ossinger_2017,Knaak_2019,Brandl_2020}. 

Determination of the spin state using STM has been shown to be possible in these measurements by analysis of the current-voltage characteristics and both mechanically- and electrically-induced switching has been reported \cite{Kuang_2017,Gopakumar_2012,Gopakumar_2013,Jasper_2017,Bairagi_2014}. Electrically-induced switching has been observed in several STM studies and can be attributed to the movement of the complex on the crystalline surface, or a true spin transition which is evidenced by a conformation change in the complex and/or the appearance of electrical signatures in tunnelling spectroscopy measurements, for example the appearance of a Kondo peak localized to the position of the metal core of the complex \cite{Miyamachi_2012,Gruber_2014}. Electrically-induced switching has also been shown to demonstrate memristor like behavior, with the resistance states selectable depending upon the magnitude and polarity of the voltage pulse~\cite{Miyamachi_2012}. Such behavior not only gives new functionalities for storing spin information on a single molecule, a key ingredient for molecular spintronic devices, but also opens the route to new strategies in spin-based neuromorphic computing.

Break junction experiments have also been used to contact individual SCO complexes and the results complement those obtained with STM. Both mechanically- and electrically-induced spin state switching has been observed in mechanically controllable break junctions (MCBJs), while electromigrated nanogap devices have also demonstrated gate-controlled spin switching \cite{Frisenda_2016, Harzmann_2015, Meded_2011}. In addition to this, spin crossover complexes coupled to graphene-based devices have been investigated in an effort to reduce surface interactions, which adversely effect the switching properties \cite{Burzuri_2018, Villalva_2021}.

Recently, the switching of a single [Fe(III)(EtOSalPet)(NCS)] SCO molecule bound to one gold electrode in a nanogap junction formed by feedback-controlled electromigration has been monitored in real time by measuring the resistance through the molecule~\cite{Gee_2020}.
At temperatures below 200\,K, two resistance levels were found and ascribed to two different spin states with $S=5/2$ and an $S=1/2$ of the molecule.

\begin{figure}
\hspace{-0.7cm}
\center
\includegraphics[width=0.9\columnwidth]{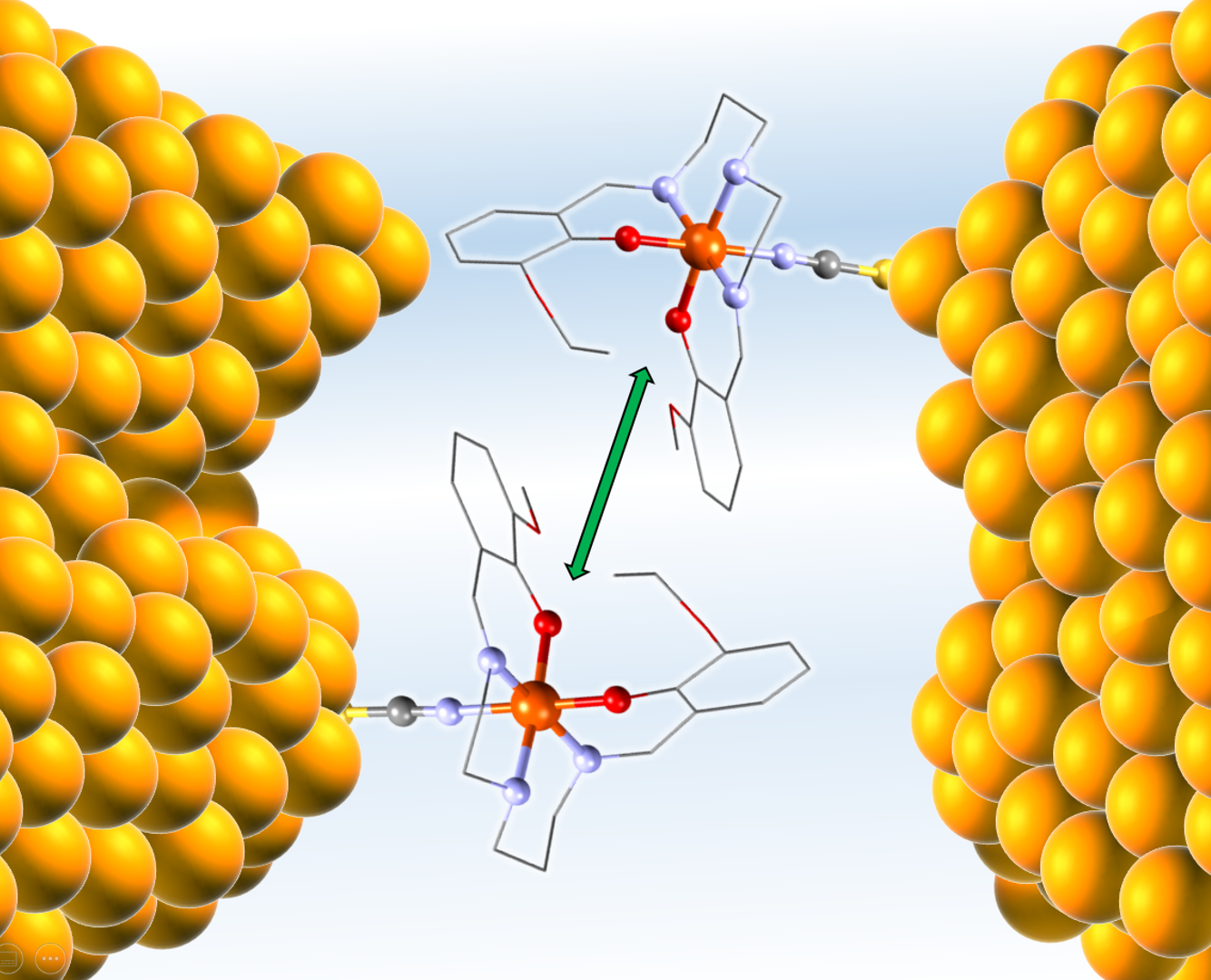}
\caption{Sketch of the studied system: Two coupled spin-crossover molecules [Fe(III)(EtOSalPet)(NCS)] residing in the gap between two gold electrodes.}
\label{fig:setup}
\end{figure}

In one sample, switching between four instead of two distinctive resistance values has been observed.
The natural interpretation is that in this sample two molecules reside within the nanogap, as illustrated in Fig.~\ref{fig:setup}.
But this immediately triggers the question whether or not the two molecules act independently of each other.
To answer this question, we perform, in this paper, a statistical analysis of the switching behavior.
We find that the two molecules indeed influence each other.

\section{\label{sec:experiment}Experimental setup}

\subsection{Device preparation and Measurement scheme}

Devices are fabricated using a combination of nanoimprint lithography and UV-lithography as described previously~\cite{Gee2_2020}. In brief, a bilayer of polymethylglutarimide (PMGI) and polymethyl methacrylate (PMMA) resist is spin coated onto an oxidized Si wafer. A stamp with the desired nanoscale features is imprinted into the heated PMMA top layer. After separation of the stamp, the residual PMMA resist is removed using oxygen plasma etching and the bottom PMGI layer is developed to produce an undercut structure. Metal is deposited onto the sample and lift-off is carried out in warm N-Methyl-2-pyrrolidone (NMP). Leads and contact pads are then connected to the constrictions using conventional UV-lithography. 

After patterning, the completed devices are wire bonded to a sample holder and a solution of the spin-crossover molecules is drop cast and allowed to evaporate. Immediately afterwards the devices are mounted into a cryostat. Feedback-controlled electromigration is then used to form a nanogap at the location of the constriction which can become bridged by one or several molecules. The resistance of the device is measured {\it in situ} after electromigration has been carried out using a lock-in technique. The lock-in reference signal (8\,mV at 989\,Hz) is applied to the device and the resistance is obtained by measuring the voltage drop across a 10$\,\Omega$ shunt resistor in series with the device. Great care was taken to ensure the switching was not induced by the applied electric field, which was set at a value that is 3 orders of magnitude smaller than experiments where field-induced conformational switching has been observed (for more detailed information see discussion in~\cite{Gee_2020}). This is also in line with theoretical predictions that critical field strengths of 1 GV\,m$^{-1}$ or more are necessary in order to electrostatically induce spin transitions~\cite{Baadji_2009}.

\subsection{Measured data}

\begin{figure*}
\centering
\includegraphics[width=1.00\textwidth]{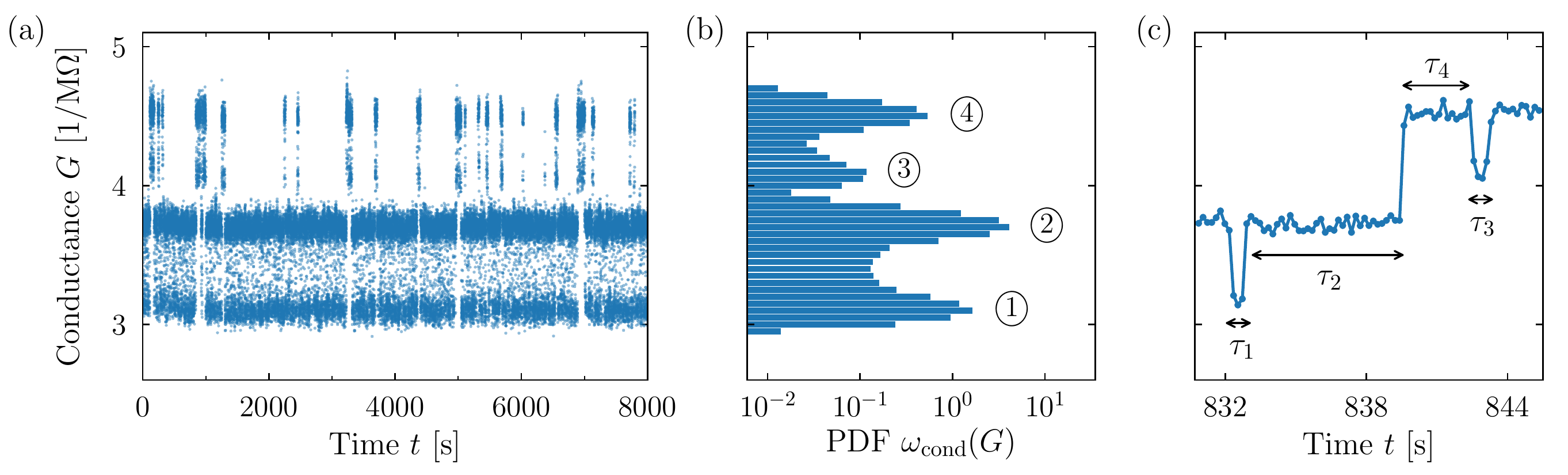}
\caption{(a) Conductance as function of time. (b) Probability density function PDF of the conductance values. Four different molecular states give rise to four maxima. (c) Segment of the conduction time trace from (a). Individual jumps between the conductance plateaus are visible. Lifetimes can be identified as indicated by $\tau_1, \tau_2, \tau_3,$ and $\tau_4$.}
\label{fig:time_trace}
\end{figure*}

The measured data is shown in Fig.~\ref{fig:time_trace}.
Panel (a) shows the conductance as a function of time for the full measurement time of 8000\,s.
The probability density function (PDF) $w_{\rm cond}(G)$ of the appearing conductance values $G$ is displayed in panel (b). 
Panel (c) depicts the conductance as a function of time for a shorter time span such that individual transitions become visible.

The distribution $w_{\rm cond}(G)$ features four peaks corresponding to four different configuration states of the system.
Each peak is characterized by its position as well as its enclosed area.
The peak position assigns a conductance value and the enclosed area an occupation probability to the respective state.

While $w_{\rm cond}(G)$ provides access to the conductance values and the occupation probabilities of the four states, an analysis of the switching behavior requires information that is not contained in the probability density function but in the full time trace of the measured conductance values.
In particular, we can determine the lifetimes of the different spin states.
Beyond that, we are able to extract the time evolution of factorial cumulants, to be defined below, which allows for an additional test of the compatibility of different models with the measured data.

\section{\label{sec:model}Model}

We assume that the switching of the conductance of the junction indicates transitions between two definite spin states (low spin and high spin) of a molecule in the junction.
Samples showing two conductance plateaus likely host only one molecule in the junction.
Here, we concentrate on the sample displaying four conductance plateaus, for which we assume two molecules to reside in the junction.
The main purpose of our analysis is to find out whether or not the two molecules act independently of each other.

\subsection{One molecule}

In the case of one molecule in the junction, we assume that the molecule is attached to one of the electrodes.
As a consequence, the current through the device is limited by tunnelling between the molecule and the second electrode.
According to the semi-classical picture of tunnelling, the tunnelling amplitude sensitively depends on the distance between molecule and second electrode.
The change from the high-spin to the low-spin state of the molecule is accompanied with a reduction of the molecule length of the order of $0.1 \rm \AA$ \cite{Ivan_2015}.
According to this picture, the low-spin (L) state corresponds to a lower conductance $G_{\tL}$, the high-spin (H) state to a higher conductance $G_{\tH}$.
The conductance change denoted by
\begin{equation}
	\Delta G = G_{\tH} - G_{\tL} > 0 
\end{equation}
can be directly read from the PDF of the conductance values as the distance between the two peaks corresponding to the high-spin and low-spin state.

The energies of the high-spin and low-spin states differ from each other. 
Whether the ground state corresponds to the high-spin or the low-spin configuration depends on specific details of how the molecule is coupled to the electrode. 
This will vary from sample to sample.
We define the excitation energy
\begin{equation}
	\Delta E = E_{\tE} - E_{\tG} > 0 
\end{equation}
where $E_{\tE}$ and $E_{\tG}$ are the energies of the excited and the ground state, respectively.

The switching between the ground and the excited state occurs due to thermal fluctuations, governed by the temperature $T$ of the environment.
The occupation probabilities $p_{\tE}$ and $p_{\tG}$ are, then, related to the excitation energy by the relation
\begin{equation}
	\frac{p_{\tE}}{p_{\tG}} = e^{-\beta\Delta E} 
\end{equation}
that is valid in thermal equilibrium.
Here, $\beta = 1/(k_\mathrm{B} T)$ is the inverse temperature.
Since the occupation probabilities of the individual states can be read from the distribution function $w_{\rm cond}(G)$ as the area under the respective peak, we can determine the excitation energy directly from $w_{\rm cond}(G)$.

From the full time trace of the measured conductance values, we can evaluate the probability distributions $w_{{\rm life}, G}(\tau)$ and $w_{{\rm life}, E}(\tau)$ of the time span $\tau$ between entering and leaving the ground and excited state, respectively.
Fitting these distributions to
\begin{equation}
	w_{{\rm life}, G}(\tau) = \frac{e^{-\tau/\tau_{\tG}}}{\tau_{\tG}}
\end{equation}
for the ground state and similarly for the excited states provides the corresponding lifetimes $\tau_{\tG}$ and $\tau_{\tE}$.

In thermal equilibrium, the detailed balance relation
\begin{equation}
	\frac{p_{\tG}}{\tau_{\tG}} = \frac{p_{\tE}}{\tau_{\tE}} 
\end{equation}
relates the lifetimes to the occupation probabilities, such that
\begin{equation}
	\frac{\tau_{\tE}}{\tau_{\tG}} = e^{-\beta\Delta E} 
\end{equation}
provides an alternative way to determine the excitation energy $\Delta E$.
Extracting $\Delta E$ in two different ways, namely from $w_{\rm cond}(G)$ on the one hand side and from $w_{{\rm life}, G}(\tau)$ as well as $w_{{\rm life}, E}(\tau)$ on the other hand side, allows for a consistency check, which increases the reliability of the determined value for the excitation energy.

\subsection{Two molecules}

The considered sample exhibits four instead of two conductance plateaus, indicating the presence of two molecules in the junction as illustrated in Fig.~\ref{fig:setup}.
From inspecting the time trace of conductance values in Fig.~\ref{fig:time_trace} with the naked eye, we immediately recognize that one molecule switches much faster than the other one
(both molecules switch, however, faster than the single-molecule device studied in Ref.~\onlinecite{Gee_2020}).
In the following, we refer to the fastly- and slowly-switching molecule as the $\tf$ and $\ts$ molecule, respectively.

Fast switching occurs between the two lowest conductance plateaus and between the two highest conductance plateaus.
Slow switching, on the other hand, takes place between the first and the third as well as between the second and the fourth conductance plateau.
Following the interpretation that the low-spin state of a molecule is associated with a reduced molecule length and, thus, a lower conductance than for the high-spin state, we are able to assign to each of the conductance peaks the corresponding spin state of the two-molecule system.
We find $G_{\tL \tL} < G_{\tH \tL} < G_{\tL \tH} < G_{\tH \tH}$, where the first index denotes the spin state of the $\tf$ and the second index the $\ts$ molecule.

We define
\begin{equation}
	\Delta G^{\tf}\big|_A = G_{\tH A} - G_{\tL A}
\end{equation}
as conductance change of the $\tf$ molecule under the condition that the $\ts$ molecule is in state $A=\tL,\tH$.
Similarly,
\begin{equation}
	\Delta G^{\ts}\big|_A = G_{A \tH} - G_{A \tL}
\end{equation}
is the conductance change of the $\ts$ molecule under the condition that the $\tf$ molecule is in state $A=\tL,\tH$.
The conductance change associated with the switching of one molecule may, in principle, depend on the state of the other one, i.e., $\Delta G^x\big|_{\tL} \neq \Delta G^x\big|_{\tH}$ for $x=\tf,\ts$.

Inspecting the probability density function $w_{\rm cond}(G)$ of the conductance values, again with the naked eye, allows us to relate for both molecules the spin to the energy state.
The occupation probability of each state is indicated by the area under the respective conductance peak.
The conductance peak at $G_{\tH \tL}$ has the largest area, i.e., the two molecules are in the ground state.
This means that for the $\tf$ molecule the ground state carries high spin, while for the $\ts$ molecule the ground state has low spin.
As mentioned earlier, the relation between spin and energy depends on details of the coupling between molecule and electrode, which, for the present sample, happens to lead to opposite results for the two molecules involved.
The full translation between spin and energy states is, then, given by $\tL \tL \leftrightarrow \tE \tG$, $\tH \tL \leftrightarrow \tG \tG$, $\tL \tH \leftrightarrow \tE \tE$, and $\tH \tH \leftrightarrow \tG \tE$.

Not only the conductance change but also the excitation energy of the $\tf$ molecule may depend on the state of the $\ts$ molecule.
Similarly as for the conductances, we define $\Delta E^{\tf}|_A$ as the excitation energy of the $\tf$ molecule under the condition that the $\ts$ molecule is in state $A=\tG, \tE$, and similarly $\Delta E^{\ts}|_A$.

Using the occupation probabilities, the excitation energies can be obtained from
\begin{align}
	\Delta E^{\tf}\big|_A &= - k_\mathrm{B} T \ln \left( \frac{p_{\tE A}}{p_{\tG A}}\right) \, 
	\\	
	\Delta E^{\ts}\big|_A &= - k_\mathrm{B} T \ln \left( \frac{p_{A \tE}}{p_{A \tG}}\right) \, 
\end{align}
In general, the excitation energy of one molecule may depend on the state of the other one, i.e., $\Delta E^x\big|_{\tE} \neq \Delta E^x\big|_{\tG}$ for $x=\tf,\ts$.

For each of the four states of the two-molecule system, we can extract the corresponding lifetime $\tau_{\tG \tG}$, $\tau_{\tG \tE}$, $\tau_{\tE \tG}$, and $\tau_{\tE \tE}$.
All these lifetimes are limited by a switching event of the $\tf$ molecule.
Therefore, we can derive the excitation energy of the $\tf$ molecule not only from the occupation probabilities but also from 
\begin{equation}
	\Delta E^{\tf}\big|_A = - k_\mathrm{B} T \ln \left( \frac{\tau_{\tE A}}{\tau_{\tG A}}\right) \, 
\end{equation}
 
To get access to the time scale governed by the $\ts$ molecule, we also determine the lifetimes $\tau^{\ts}_{\tG}$ and $\tau^{\ts}_{\tE}$ of the ground and the excited state of the $\ts$ molecule alone by ignoring switching events of the $\tf$ molecule. 

\begin{figure*}[t]
\centering
\includegraphics[width=1.00\textwidth]{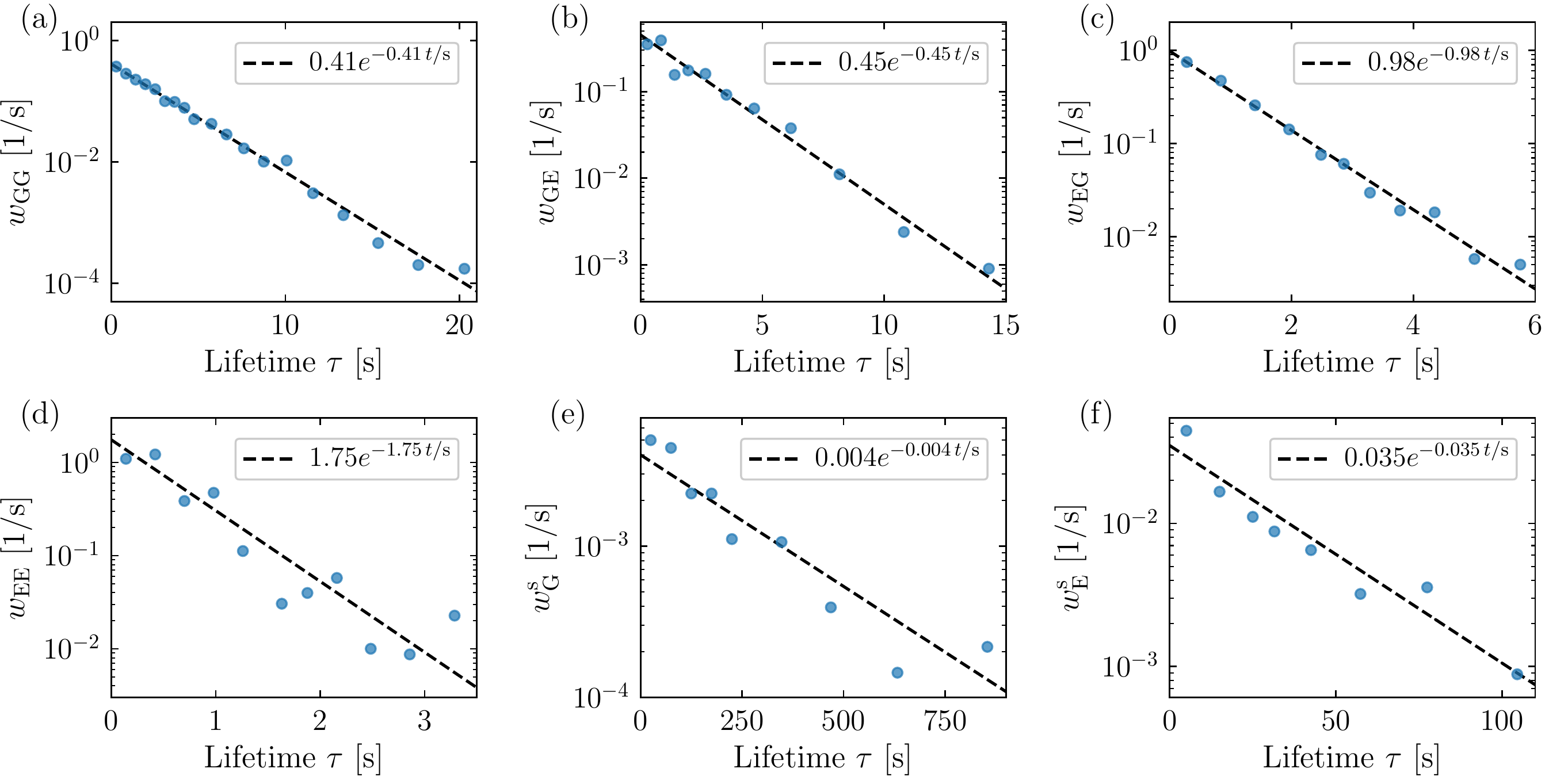}
\caption{Lifetime distributions of the different states of the two-molecule system and the lowly switching molecule.}
\label{fig:life_time}
\end{figure*}

\subsection{Independent vs coupled molecules}

Suppose, the two molecules act independently of each other.
Then, the total conductance through the junction should be just the sum of the conductances through the two molecules forming a parallel circuit.
In addition, conductance through the $\tf$ molecule should not depend on the state of the $\ts$ molecule and vice versa.
This implies
\begin{equation}
 	\Delta G^x|_{\tL} = \Delta G^x|_{\tH}
\end{equation}
for $x=\tf,\ts$.
Similarly, the excitation energy of one of the molecules should be independent of the state of the other molecule, 
\begin{equation}
	\Delta E^x|_{\tL} = \Delta E^x|_{\tH}
\end{equation}
for $x=\tf,\ts$.
Furthermore, the lifetime of any two-molecule state $AB$ with $A=\tG,\tE$ and $B=\tG,\tE$ should be decomposable into
\begin{equation}
	\frac{1}{\tau_{AB}} = \frac{1}{\tau^{\tf}_{A}} + \frac{1}{\tau^{\ts}_{B}}
\end{equation}
where $\tau^x_A$ is the lifetime of the $A$ state of the $x$ molecule alone. 
Since in our sample, the lifetime of the $\ts$ molecule is much larger than that of the $\tf$ molecule, independence of the two molecules implies
\begin{equation}
	\tau_{A\tG} = \tau_{A\tE}
\end{equation}
for $A=\tG,\tE$.

Any deviation from these equalities indicates the presence of interaction between the two molecules.
In our statistical analysis below, we will find that the behavior of all three quantities, the conductances, the excitation energies, and the lifetimes are incompatible with the assumption of two independent molecules.

\section{\label{sec:analysis}Statistical analysis}

We identify the values of the conductance plateaus by the local maxima of the probability density~ $w_{\rm cond}(G)$.
The occupation probabilities are obtained by integrating the probability density $w_{\rm cond}(G)$ around the four peaks.
Thereby, we choose the local minima between the peaks as the upper/lower end of the integration range.
To get the lifetimes of the four states, we fit an exponential decay to the life-time distribution functions, as shown in Fig.~\ref{fig:life_time}.
The results are summarized in Table~\ref{tab1}.
We now check whether or not these values are compatible with the model of independent molecules.

\begin{table}
	\begin{tabular}{|c||c|c|c|c|}
	\hline
	& $ \circled{1} $ & $\circled{2}$ & $\circled{3}$ & $\circled{4}$ \\
	\hline\hline
	spin state & \tL \tL & \tH \tL & \tL \tH & \tH \tH \\
	\hline
	energy state & \tE \tG & \tG \tG & \tE \tE & \tG \tE \\
	\hline
	conductance $G \left[ {\rm M\Omega}^{-1}\right]$ & 3.11 & 3.71 & 4.08 & 4.51 \\
	\hline
	occupation probability $p\left[ \% \right]$ & 26.2 & 63.1 & 2.3 & 8.4 \\
	\hline
	lifetime $\tau \left[ {\rm s}\right]$ & 1.02 & 2.44 & 0.57 & 2.22 \\
	\hline
	$\ts$-molecule lifetime $\tau^{\ts} \left[ {\rm s}\right]$ & \multicolumn{2}{c|}{250} &  \multicolumn{2}{c|}{29} \\
	\hline
	\end{tabular}
\caption{\label{tab1}
Results from the statistical analysis:
the first index refers to the $\tf$, the second to the $\ts$ molecule.
}
\end{table}

\subsection{Conductance}

We find that the conductance changes $\Delta G^{\tf}|_{\tL}=0.60 \,{\rm M \Omega}^{-1}$ and $\Delta G^{\tf}|_{\tH}=0.43\, {\rm M \Omega}^{-1}$ for the $\tf$ molecule depend on the state of the $\ts$ molecule and vice versa, $\Delta G^{\ts}|_{\tL}=0.97\, {\rm M \Omega}^{-1}$ and $\Delta G^{\ts}|_{\tH}=0.80\, {\rm M \Omega}^{-1}$.
In conclusion, 
\begin{equation}
	\Delta G^{\tf}\big|_{\tL} - \Delta G^{\tf}\big|_{\tH} = \Delta G^{\ts}\big|_{\tL} - \Delta G^{\ts}\big|_{\tH} = 0.17\, {\rm M \Omega}^{-1} 
\end{equation}
is significantly different from zero, indicating interaction between the two molecules.
The conductance change of one molecule is larger when the other molecule is in the low-spin state as compared to the case when the other molecule is in the high-spin state.

\subsection{Excitation energy}

From the occupation probabilities of the four states, we derive the excitation energies (in units of the energy scale set by temperature) as $\Delta E^{\tf}|_{\tG}=0.88\, k_\mathrm{B}T$, $\Delta E^{\tf}|_{\tE}=1.30\, k_\mathrm{B}T$, $\Delta E^{\ts}|_{\tG}=2.01\, k_\mathrm{B}T$, and $\Delta E^{\ts}|_{\tE}=2.43\, k_\mathrm{B}T$.
In conclusion, we find
\begin{equation}
	\Delta E^{\tf}\big|_{\tE} - \Delta E^{\tf}\big|_{\tG} = \Delta E^{\ts}\big|_{\tE} - \Delta E^{\ts}\big|_{\tG} = 0.42\, k_\mathrm{B}T \, 
\end{equation}
i.e., the excitation energy of one molecule is significantly larger when the other molecule is in the excited state as compared to the case when the other molecule is in the ground state.
This, again, indicates a coupling of the two molecules.

As an independent check, we also calculate the excitation energies of the $\tf$ molecules via the lifetimes. 
This yields $\Delta E^{\tf}|_{\tG}= 0.87\, k_\mathrm{B}T$ and $\Delta E^{\tf}|_{\tE}= 1.36\, k_\mathrm{B}T$, in good agreement with the values obtained from the occupation probabilities.

\subsection{Lifetime}

The lifetimes $\tau_{AB}$ with $A=\tG,\tE$ and $B=\tG,\tE$ are dominated by the switching of the $\tf$ molecule.
For independent molecules, we expect $\tau_{A\tG}=\tau_{A\tE}$ for $A=\tG,\tE$.
From our data, however, we find
\begin{align}
	\tau_{\tE \tG} - \tau_{\tE \tE} &= 0.45\, {\rm s}\, 
	\\
	\tau_{\tG \tG} - \tau_{\tG \tE} &= 0.22\, {\rm s} \, 
\end{align}
i.e., the $\tf$ molecule switches more often when the $\ts$ molecule is in the ground/low-spin state as compared to when the $\ts$ molecule is in the excited/high-spin state.

\subsection{Time-dependent full counting statistics}

In the above analysis, we have characterized the switching behavior of the molecules and found evidence for a coupled two-molecule system rather than independent switching of two individual molecules.
To reach this conclusion, however, we only used a small part of the information that is contained in the measured time trace of conductance values.
To achieve a more complete picture, we now analyze the full counting statistics of the switching events in more detail.

\begin{figure}
\hspace{-0.7cm}
\includegraphics[width=0.93\columnwidth]{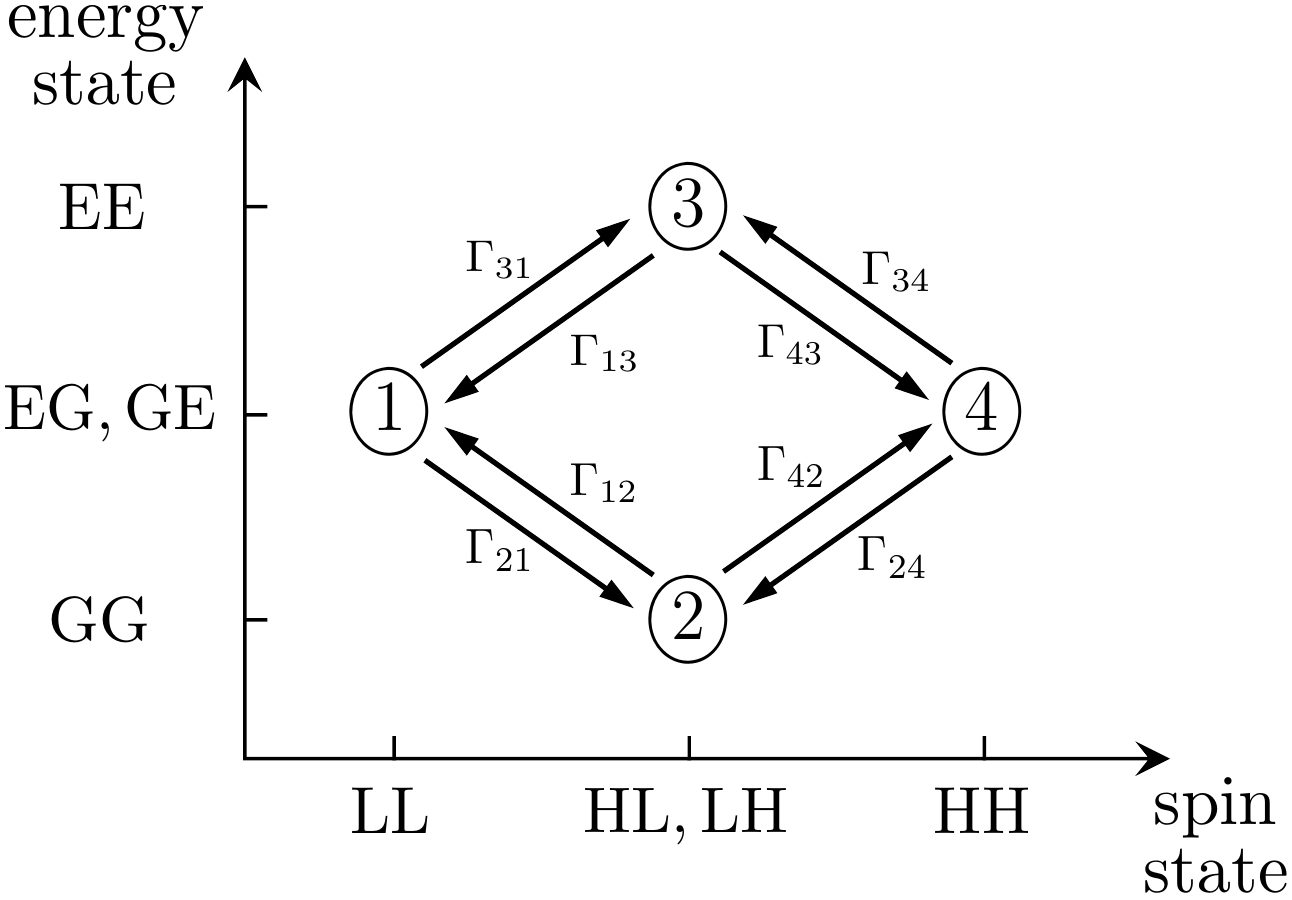}
\caption{Stochastic system of the two molecules: the four possible states and connecting transitions.}
\label{fig:stochastic_system}
\end{figure}

\begin{figure*}[t]
\centering
\includegraphics[width=1.00\textwidth]{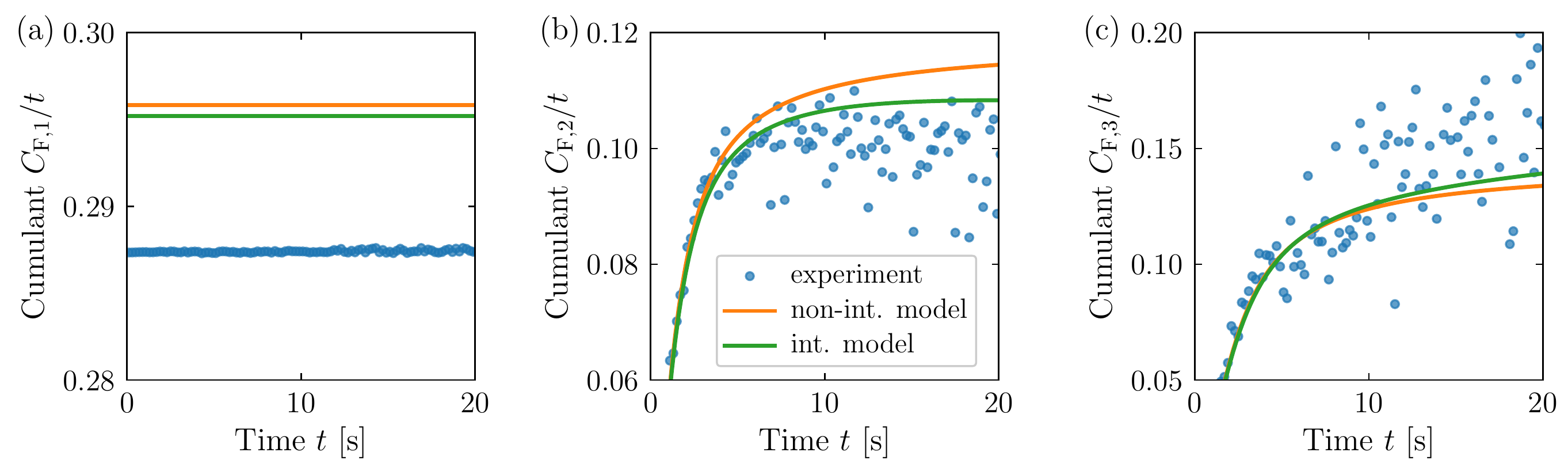}
\caption{Factorial cumulants probing the switching dynamics of the two molecules. Experimental data is compared with the theoretical model of non-interacting and interacting molecules.}
\label{fig:cumulants}
\end{figure*}

The two-molecule system is described by the following stochastic system, sketched in Fig.~\ref{fig:stochastic_system}.
We label the four state of the two-molecule system by $\chi=1,2,3,4$.
The corresponding spin or energy state can be read from Table~\ref{tab1}.
For example, $\circled{1}$ refers to spin state $\tL \tL$ and energy state $\tE \tG$.
The transition rate from state $\chi$ to $\chi'$ is denoted by $\Gamma_{\chi' \chi}$.
In total, there are eight transitions in which one of the two molecules switches.
However, since the $\tf$ molecule switches much faster than the $\ts$ molecule, the switching rate of the $\ts$ molecule can be considered as independent of the state of the $\tf$ molecule, i.e., there are only six independent transition rates.
They can be determined by the measured lifetimes.
We use $\Gamma_{12}+\Gamma_{42}=1/\tau_{\tG \tG}$, $\Gamma_{21}+\Gamma_{31}=1/\tau_{\tE \tG}$,  $\Gamma_{24}+\Gamma_{34}=1/\tau_{\tG \tE}$, $\Gamma_{13}+\Gamma_{43}=1/\tau_{\tE \tE}$, as well as $\Gamma_{31}=\Gamma_{42}=1/\tau^{\ts}_{\tG}$ and $\Gamma_{13}=\Gamma_{24}=1/\tau^{\ts}_{\tE}$, with the values of the lifetimes as listed in Table~\ref{tab1}.
With this information, the stochastic system for the coupled two-molecule system is fully defined.

In the following, we want to compare the interacting two-molecule system with the simpler model of two non-interacting molecules, for which the switching rate of the $\tf$ molecule is assumed to be independent of the state of the $\ts$ molecule, $\tau_{A\tG}=\tau_{A\tE}$ with $A=\tG,\tE$.
Therefore, we replace for the non-interacting model the rates associated with switching of the $\tf$ molecule by the averages 
\begin{align}
	\overline \Gamma_{12} = \overline \Gamma_{34}  =\, \, &\frac{ p_{\tG \tG}\Gamma_{12} + p_{\tG \tE} \Gamma_{34}}{p_{\tG \tG}+p_{\tG \tE}}\,
	\\
	\overline \Gamma_{21} = \overline \Gamma_{43} = \, \, & \frac{p_{\tE \tG}\Gamma_{21} +p_{\tE \tE}\Gamma_{43}}{p_{\tE \tG}+p_{\tE \tE}}\,
\end{align}
with the values of the lifetimes and occupation probabilities as listed in Table~\ref{tab1}.

To analyze the full counting statistics of the switching events, we first need to decide what to count.
We choose to count the events of exciting one of the two molecules but our results remain qualitatively the same if we choose to count the events of relaxation.
Moreover, they remain the same if we count events either increasing or decreasing the spin, which indicates that detailed balance is fulfilled~\cite{Stegmann_2017}.
The entire time trace of measured conductance values is divided into slices of length $t$.
For each time slice, we count the number $N$ (with $N\ge 0$) of events of exciting one of the molecules.
Averaging over all time slices yields the probability distribution $P_N(t)$, referred to as the full counting statistics.

A  distribution function is conveniently characterized by its moments or cumulants.
For the integer-valued stochastic variable $N$, it is advantageous to employ so-called factorial cumulants $C_{\text{F},m}(t)$ obtained from the $m$-th derivative
\begin{equation}
	C_{\rm F,m}(t) = \partial_z^m \ln \mathcal{M}_{\rm F}(z,t)\big|_{z=0}
\end{equation}
of the generating function
\begin{equation}
	\mathcal{M}_{\rm F}(z,t) = \sum_N (z+1)^N P_N(t) 
\end{equation}
which is the (shifted) $z$-transform of the probability function $P_N(t)$.

There are several reasons why factorial cumulants are better suited than ordinary cumulants for the analysis of stochastic systems with integer-valued variables.
First, ordinary cumulants generically change sign as a function of time \cite{Flindt_2009,Fricke_2010}, whereas the factorial cumulants of the very same probability distribution are much more well-behaved \cite{Komijani_2013}.
In fact, it can be shown that sign changes of the factorial cumulants indicate the presence of interaction-induced correlations \cite{Kambly_2011,Stegmann_2015,Stegmann_2016}.
Second, factorial cumulants naturally appear in counting bosons or fermions as a consequence of normal ordering of the field operators \cite{Koenig_2021}.
Third, factorial cumulants are less sensitive to systematic errors introduced by the detector, such as a limited time resolution and noise on the detector signal, as well as to statistical errors due the finite length of the measured time trace \cite{Kleinherbers_2021}.
These advantages made it possible, e.g., to determine the spin-relaxation rate of Zeeman-split quantum-dot levels in an optical detection of charge transfers \cite{Kurzmann_2019} and to identify the delocalized and coherent nature of antiferromagnetic spin excitations of Mn$_4$ complexes coupled to carbon nanotubes from the time-dependent electric current \cite{Besson_2021}.
Factorial cumulants have also been suggested to identify violation of detailed balance~\cite{Stegmann_2017},  hidden states~\cite{Stegmann2_2017} coherent dynamics~\cite{Stegmann_2018} and attractive electron-electron interaction~\cite{Kleinherbers_2018} in quantum-dot systems. 

In Fig.~\ref{fig:cumulants}, we show the first three factorial factorial cumulants of the switching probability distribution as a function of time. 
We compare the experimental data with two model calculations, one for a non-interacting and one for the interacting model.
The theoretical curves are obtained from the generating function
\begin{equation}
	\mathcal{M}_{\rm F}(z,t) = (1,1,1,1) \cdot \exp( \mathbf{W}_{\! z+1} t)  \mathbf{p}_\mathrm{stat}
\end{equation}
where
\begin{equation}
\begin{footnotesize}
	\mathbf{W}_{\!z} = \left( 
	\begin{array}{cccc} 
	-\Gamma_{21}-\Gamma_{31} & z \Gamma_{12} & \Gamma_{13} & 0 \\
	\Gamma_{21} & -\Gamma_{12}-\Gamma_{42} & 0 & \Gamma_{24} \\
	z \Gamma_{31} & 0 & -\Gamma_{13}-\Gamma_{43} & z \Gamma_{34} \\
	0 & z \Gamma_{42} & \Gamma_{43} & -\Gamma_{24}-\Gamma_{34} 	
	\end{array} \right)
	\end{footnotesize}
\end{equation}
is the transition matrix and the vector $\mathbf{p}_\mathrm{stat}$ of the stationary probability distribution is obtained from $\mathbf{W}_{\! 1}\mathbf{p}_\mathrm{stat}=\mathbf{0}$.

We find reasonable agreement of the measured first and third factorial cumulant (blue points in Figs.~\ref{fig:cumulants}a,c) with the non-interacting (orange line) and interacting model (green line) of two coupled molecules. However, the second factorial cumulant depicted in~\ref{fig:cumulants}b is reproduced more precisely by the interacting model.

\section{\label{sec:con}Conclusions}

We analyzed the switching behavior of two spin-crossover molecules residing a nanojunction.
We determined the resistance values, the occupation probabilities, and the lifetimes of the spin states of the two molecules and find striking evidence that they are coupled to each other.
This analysis is complemented by studying the time dependence of the distribution of the number of switching events by making use of factorial cumulants.
The result supports the conclusion that the measured data are incompatible with the assumption of two independently-switching spin-crossover molecules.

\begin{acknowledgments}
We thank M. Gruber for useful discussions.
We acknowledge financial support from the Deutsche Forschungsgemeinschaft (DFG, German Research Foundation) under Project-ID 278162697 - SFB 1242. PS acknowledges support from the German National Academy of Sciences Leopoldina (Grant No.~LPDS 2019-10).
\end{acknowledgments}


\end{document}